\documentclass[reprint, superscriptaddress, amsmath, amssymb, aps, prl, floatfix]{revtex4-2}

\usepackage{graphicx}
\usepackage{dcolumn}
\usepackage{bm}
\usepackage{commath}
\usepackage{physics}

\begin{document}

\preprint{APS/123-QED}

\title{Impact of Monoatomic Vacancies in 2D Materials on the Performance of Magnetic Tunnel Junction Devices: Insights from Configurations and Interface Interactions}

\author{Halimah Harfah} 
\email{Corresponding author: harfah.h@opt.mp.es.osaka-u.ac.jp}
\thanks{These authors contributed equally to this work}
\affiliation{Graduate School of Engineering Science, Osaka University, 1-3 Machikaneyama-Cho, Toyonaka, Osaka 560-0043, Japan }

\author{Yusuf Wicaksono}
\thanks{These authors contributed equally to this work}
\affiliation{RIKEN Cluster for Pioneering Research (CPR), 2-1 Hirosawa, Wako, Saitama 351-0198, Japan}

\author{Gagus Ketut Sunnardianto}
\affiliation{
Research Center for Quantum Physics, National Research and Innovation Agency of Indonesia (BRIN), Tangerang Selatan, Banten, 15314, Indonesia
}
\affiliation{Research Collaboration Center for Quantum Technology 2.0, National Research and Innovation Agency of Indonesia (BRIN), Bandung 40132, Indonesia}
\affiliation{School of Materials Science and Engineering, Nanyang Technological University, 50 Nanyang Avenue, Singapore 639798, Singapore}

\author{Muhammad Aziz Majidi}
\affiliation{
 Department of Physics, Faculty of Mathematics and Natural Sciences, Universitas Indonesia, Kampus UI Depok, Depok 16424, Indonesia.
}

\author{Koichi Kusakabe}%
\affiliation{
 School of Science, Graduate School of Science, University of Hyogo 3-2-1 Kouto, Kamigori-cho, Ako-gun, 678-1297, Hyogo, Japan.
}

\date{\today}

\begin{abstract}
We investigate the impact of monoatomic vacancies in 2D materials on the performance of magnetic tunnel junction (MTJ) devices using first-principles calculations within Density Functional Theory (DFT). Specifically, we analyze the influence on hexagonal boron nitride (hBN) with various layer configurations, uncovering distinct transmission probability patterns. Transmission calculations were conducted using the Landauer-Büttiker formula employing the Non-Equilibrium Green's Function (NEGF) method.
In the Ni/hBN(V$_B$)-hBN/Ni system, a significant reduction in transmission probability was observed compared to non-vacancy configurations. However, when two hBN vacancies were considered, creating the Ni/hBN(V$_B$)-hBN(V$_B$)/Ni MTJ system, a new transmission channel mediated by vacancy localized states emerged. The introduction of a monoatomic boron vacancy in the middle hBN layer of the Ni/3hBN/Ni system revealed nuanced effects on the transmission probability, highlighting alterations in the spin minority and majority channels.
Additionally, we explore the monoatomic vacancy in the graphene layer in the Ni/hBN-Gr-hBN/Ni MTJ, uncovering a unique transmission channel influenced by the proximity effect. Our findings suggest that the creation of monoatomic vacancies on the insulator barrier of 2D materials induces distinctive characteristics shaped by the interaction between the surface state of the electrode and the localized state of the monoatomic vacancy layer in the MTJ system.
\end{abstract}

\maketitle

\section{Introduction}

Magnetic tunnel junctions (MTJs) have attracted significant attention in the realm of spintronics, owing to their diverse array of potential applications ranging from logic devices to magnetic sensors \cite{Chappert:2007,Dery:2007,Zhu:2006,Childress:2005,Chen:2010,Iqbal:2018-review}. A critical factor in optimizing MTJ performance is the tunneling magnetoresistance (TMR) ratio, wherein MgO has emerged as the predominant choice for the tunnel barrier material. For instance, the CoFeB/MgO/CoFeB MTJ configuration has showcased remarkable TMR values, reaching as high as 1100\% at 4.2 K \cite{ikeda:2010}. However, the pursuit of downsizing these devices by reducing the barrier thickness has often been impeded by a concomitant decrease in TMR, sometimes plummeting to 55\%, primarily attributed to the presence of uncontrollable defects within the MgO tunnel barrier \cite{Reiss2013:thin-film_MgO,Wang2011}.

Conversely, researchers have been actively exploring alternatives to MgO, seeking to replace it with 2D materials as a means of device miniaturization. The potential of 2D materials such as graphene (Gr) or hexagonal boron nitride (hBN) as MTJ spin valves has been thoroughly investigated \cite{yusuf1,yusuf2,yusuf3,Harfah2020,Harfah2022,hBN-dev-2022,hBN-MTJ,Harfah2024}. One such approach involves employing a monolayer or a few layers of hBN as the tunnel barrier, sandwiched between ferromagnetic electrodes to form MTJ structures \cite{Piquemal_Banci:2017,Exfolation1,Asshoff:2017,Iqbal:2015-JMCC,Iqbal:2013-NanoRes,Chen:2013,Li:2014,LiWan:2014,Mandal:2012,Martin:2015,Entani:2016}. In the Ni/monolayer hBN/Ni MTJ system, the interaction between Ni and N atoms at the interface can lead to $pd$-hybridization, allowing electrons at the Fermi energy to easily pass through the monolayer hBN as a tunnel barrier \cite{Harfah2020}. This electron transmission is primarily due to propagating-wave electrons. In contrast, when multilayer hBN serves as the tunnel barrier, the transmission of electrons through the hBN insulator barrier is facilitated by the Ni d$_{z^2}$-orbital surface state and the proximity effect induced on the hBN layer \cite{Harfah2022}. This finding emphasizes the pivotal role played by the proximity effect within 2D materials in modulating electron tunneling capacity. Interestingly, unlike conventional thin-film materials such as MgO, modified 2D materials, engineered through the introduction of monoatomic vacancies, exhibit a spectrum of unique physical and chemical properties \cite{hBN-VB1,hBN-VB2,hBN-VB3,hBN-VB4}. These distinct characteristics may mediate novel transmission phenomena not observed in MgO-based MTJs.

In the field of spintronics, investigating the impact of monoatomic vacancies in two-dimensional (2D) materials on the performance of MTJ devices is of critical importance \cite{asshoff2018magnetoresistance,ahn20202d,yu2022tunable, taudul2020impact}. To understand how single-atom defects (monoatomic vacancies) in two-dimensional (2D) materials affect the transmission properties of MTJs, we use a theoretical framework based on Density Functional Theory (DFT). This allows us to explore the complex relationship between the electronic properties and device performance. We examine the transmission properties using the Landauer-Büttiker formula. The Non-Equilibrium Green's Function (NEGF) method is then employed to gain insights into the transport properties of the systems. We focused on hexagonal boron nitride (hBN), a well-known insulator, and explored how different layer arrangements within the MTJ device affect its performance. By controlling monoatomic vacancies and altering the layer structures, we observed significant changes in the transmission probability, especially in the spin-dependent channels. Furthermore, we incorporated graphene layers within the MTJ design and investigated how monoatomic vacancies impact the transmission properties in these hybrid systems. This study deepens our understanding of the fundamental mechanisms governing electron transport in MTJ devices and provides practical guidance for designing and optimizing future spintronic devices. By investigating the impact of monoatomic vacancies in 2D materials on device performance, we aim to advance spintronics towards devices with enhanced functionality and efficiency.

\section{Computational Method}

\begin{figure}[tb]
\centering
\includegraphics[width=\columnwidth]{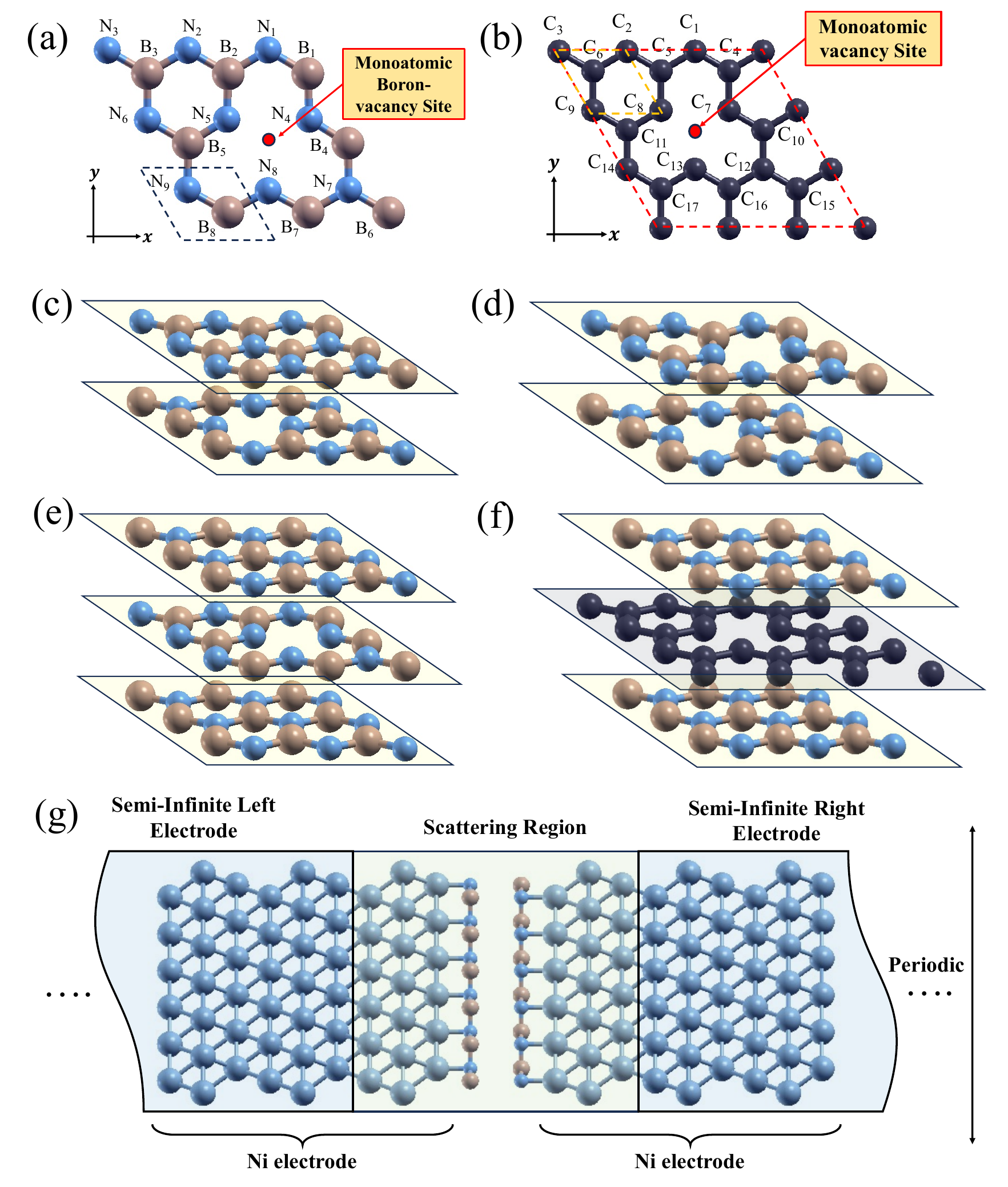}
\caption{\label{fig:method_4} A $3 \times 3$ unit cell of (a) hBN with a monoatomic boron vacancy (V$_B$), (b) graphene (Gr) with a vacancy, (c) hBN(V$_B$)/hBN, (d) hBN(V$_B$)/hBN(V$_B$), (e) hBN/hBN(V$_B$)/hBN, and (f) hBN/Gr(V)/hBN used in the calculations. Blue, black, and pink color balls represent N, C, B atoms, respectively. (g) The setup to calculate the tunneling transmission probability of the aforementioned tunnel barrier where Ni is used as electrodes. The scattering region comprises a three-layer Ni/tunnel barrier/three-layer Ni, and the left and right electrodes comprise six Ni layers. Visualization was performed using XCrySDen \cite{xcrysden}}
\end{figure}

In this work, we used a $3 \times 3$ supercell to model the MTJ systems. To introduce defects, a monoatomic boron vacancy (V$_B$) was deliberately created within the hBN layer, as depicted in Figure \ref{fig:method_4}(a). Additionally, a monoatomic vacancy was introduced into the graphene layer, as illustrated in Figure \ref{fig:method_4}(b). We explored various stacking configurations of hBN(V$_B$) layers, including hBN(V$_B$)/hBN, hBN(V$_B$)/hBN(V$_B$), and hBN/hBN(V$_B$)/hBN, as shown in Figures \ref{fig:method_4}(c), (d), and (e), respectively. Furthermore, we investigated the hBN/Gr/hBN stacking with different vacancy configurations, where the vacancy was positioned on the graphene layer, as depicted in Figure \ref{fig:method_4}(f). The study primarily focused on evaluating the significance of localized states in spin-dependent electron tunneling between two ferromagnetic electrodes, with nickel chosen as the electrode material. Given that the anti-parallel configuration (APC) is known to exhibit low transmission probability due to the spin-blocking effect, our analysis primarily focused on the parallel configuration (PC) state of the MTJ, where the magnetic moments of the upper and lower Ni electrodes were oriented parallelly upward.

The SIESTA package \cite{siesta1,siesta2} was utilized for various calculations, encompassing structural equilibrium determination, assessment of magnetic properties, mapping of spin-charge density, and determination of the local density of states (LDOS) utilizing spin-polarized DFT. Electron-ion interaction within the generalized gradient approximation (GGA) was modeled using the Troullier--Martins \cite{tmpseudo} pseudopotential and the Perdew--Burke--Ernzerhof functional, specifically the PBESol functional \cite{PBEsol}. A basis set incorporating double-zeta and polarization was employed \cite{dzp1,dzp2,dzp3}. The atomic positions were relaxed with a force tolerance of 0.01 eV/\AA, and a Monkhorst--Pack k-mesh of $36 \times 36 \times 1$ was utilized for calculations, with a mesh cutoff of 500 Ry. Additionally, van der Waals interactions between hBN and graphene were accounted for by integrating a Grimme-type dispersion potential \cite{grimme}.

The tunneling transmission probability was computed using the Landauer-Büttiker formalism within the NEGF method. The setup for calculating the transmission probability is depicted in Figure \ref{fig:method_4}(g). The spin-dependent current was determined utilizing the Landauer-Büttiker equation:

\begin{equation}\label{eq:LB_method}
I^{\uparrow(\downarrow)} = \frac{e}{h} \int_{-\infty}^\infty T^{\uparrow(\downarrow)}(E)\Big[f_L(E,\mu) - f_R(E,\mu)\Big]dE,
\end{equation}

\noindent where $f_L(E,\mu)\big(f_R(E,\mu)\big)$ represents the right (left) moving electrons injected from the left (right) electrode, and $\mu_L\big(\mu_R\big)$ denotes the chemical potentials of the left (right) electrodes, both assumed to be at the Fermi level ($E_F$) due to the zero bias voltage. The transmission probability, $T$, as a function of energy, $E$, is described by the Green's function:

\begin{equation}\label{eq:T-matrix_NEGF}
T^{\uparrow(\downarrow)}(E) = \Tr{\Big[\Gamma_LG^R\Gamma_RG^A\Big]},
\end{equation}

\noindent where $\Gamma_L(\Gamma_R)$ represents the coupling matrix of the left (right) electrode, and $G^R(G^A)$ denotes the retarded (advanced) Green's function of the central region.

\section{Results and Discussion}

\subsection{Influence of Monoatomic Boron Vacancy in One or Two hBN in Ni/2hBN/Ni MTJ}

\begin{figure}[b]
\centering
\includegraphics[width=\columnwidth]{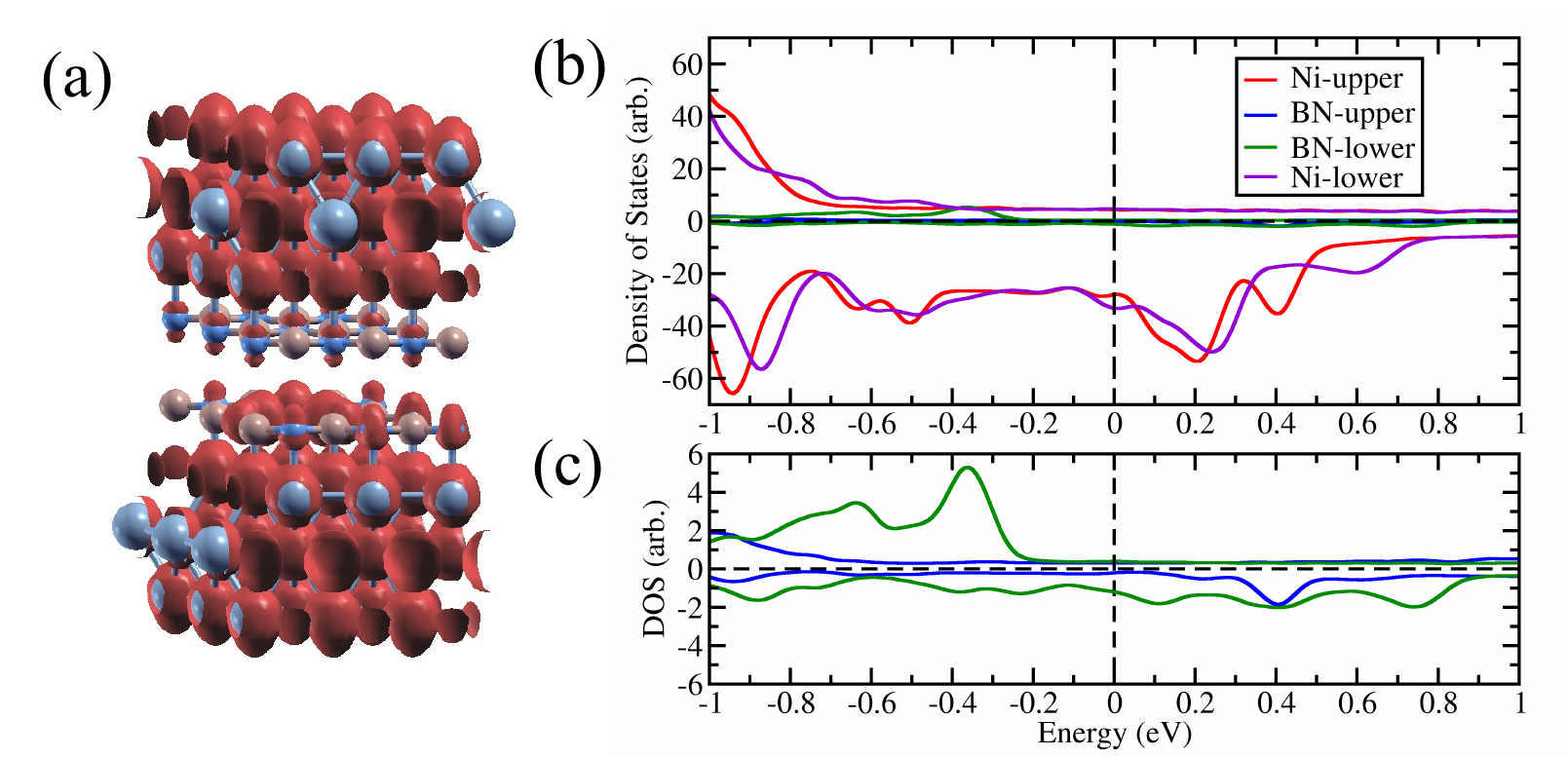}
\caption{\label{fig:3rd_figure1} (a) The $3 \times 3$ Ni/hBN/hBN(V$_B$)/Ni SCDM (red color represents spin-up electron density). Visualization was performed using XCrySDen \cite{xcrysden}. (b) LDOS for Ni-upper, hBN-upper, hBN-lower (hBN with V$_B$), and Ni-lower layers in PC configuration. The positive (negative) value of DOS represents the spin majority (minority) channel. (c) A zoom on the lower energy of LDOS in (b).}
\end{figure}

\begin{figure}[tb]
\centering
\includegraphics[width=\columnwidth]{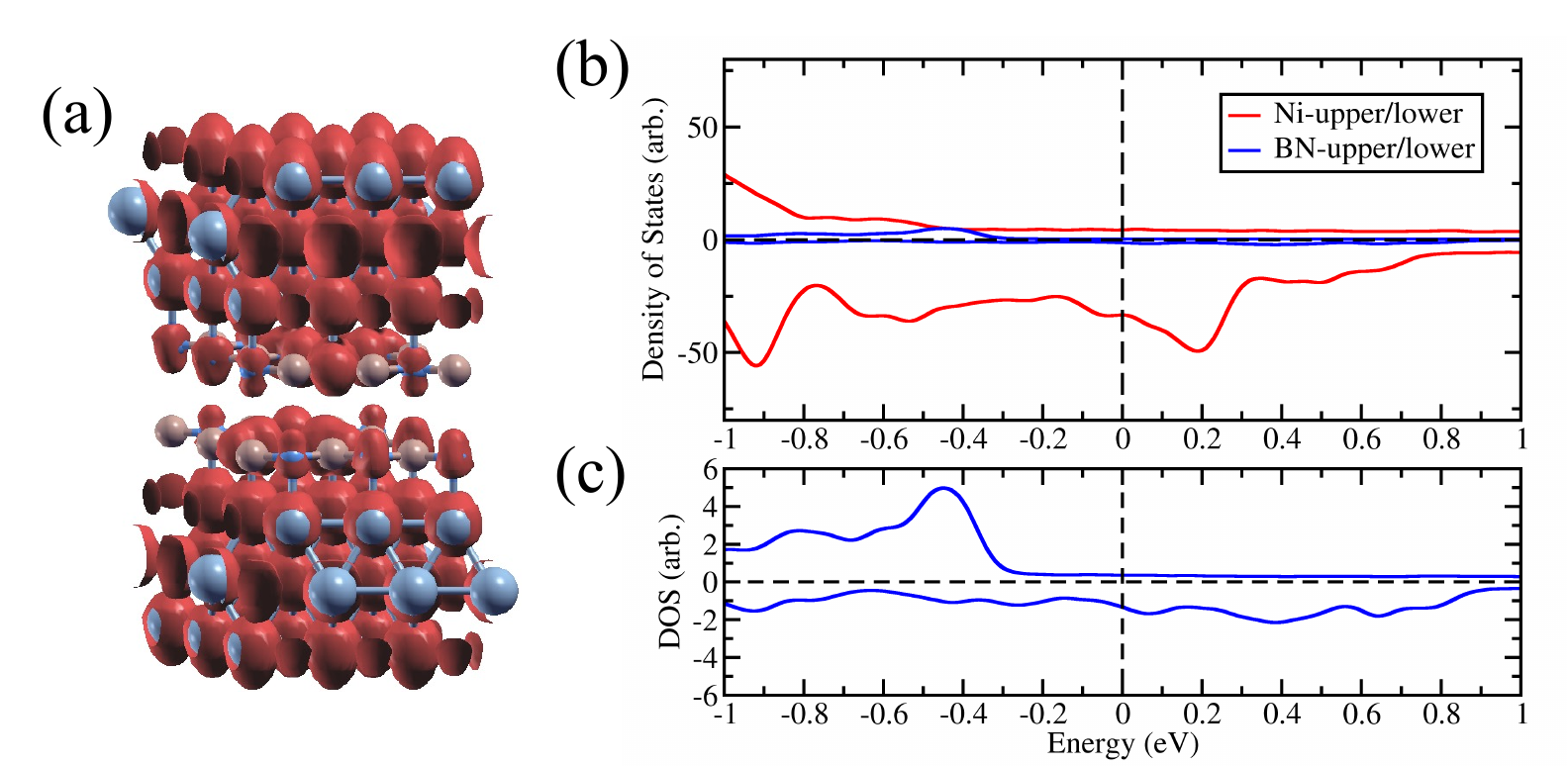}
\caption{\label{fig:3rd_figure2} (a) The $3 \times 3$ Ni/hBN(V$_B$)/hBN(V$_B$)/Ni SCDM (red color represents spin-up electron density). Visualization was performed using XCrySDen \cite{xcrysden}. (b) LDOS for Ni-upper/lower and hBN-upper/lower (which both have V$_B$) in the PC configuration. The positive (negative) value of DOS represents the spin majority (minority) channel. (c) A zoom on the lower energy of LDOS in (b). }
\end{figure}

To examine the impact of a monoatomic boron vacancy in hBN on the functionality of an MTJ device, we investigated two potential configurations: Ni/hBN(V$_B$)-hBN/Ni and Ni/hBN(V$_B$)-hBN(V$_B$)/Ni MTJ systems. The presence of a monoatomic boron vacancy induces magnetization in the hBN layer. This phenomenon is attributed to the unpaired electron of the nitrogen atoms at the vacancy site, which fails to form a $\sigma$-bond with the boron atom. We observe vacancy-induced magnetization in both Ni/hBN(V$_B$)-hBN/Ni and Ni/hBN(V$_B$)-hBN(V$_B$)/Ni MTJs, as depicted in Figures \ref{fig:3rd_figure1}(a) and \ref{fig:3rd_figure2}(a), respectively, through spin charge density mapping (SCDM). Notably, the SCDM reveals dominant spin-up electron density in the localized state near V$_B$, aligned parallel to the spin-up magnetization direction of the Ni slabs. This alignment signifies a strong ferromagnetic coupling between the magnetic moment at the vacancy and the Ni slabs. Furthermore, a short-range magnetic interaction between the localized state and neighboring nitrogen atoms is evident from the damping of spin-up electron density amplitude. This damping occurs in regions where spin-up electron density concentrates near the vacancy but decreases for nitrogen atoms farther away. The emergence of vacancy-induced magnetization in hBN(V$B$) is expected to alter its electronic structure and the propagation of Ni d${z^2}$-orbital surface state through the insulator barrier of two hBN layers.

Initially, we examined the influence of V$_B$ on the electronic structure of the MTJ system using Ni/hBN(V$_B$)-hBN/Ni. This configuration allowed for a direct comparison between pristine hBN, hybridized with the upper Ni slab, and hBN(V$_B$), hybridized with the lower Ni slab. Figure \ref{fig:3rd_figure1}(b) displays the local density of states (LDOS) of the upper and lower Ni slabs, pristine upper hBN, and lower hBN with V$_B$ in the PC configuration. A notable difference in LDOS between the upper and lower Ni slabs is observed, stemming from the modification of $pd$-hybridization at the interface of hBN(V$_B$)/Ni. This modification is evidenced by the DOS peak at $E-E_F=0.4$ eV in the spin minority channel, corresponding to hybridization at the Ni/hBN interface between the Ni d${z^2}$ orbital and the nitrogen p$_z$ orbital. This peak also appears in the spin majority channel at approximately $E-E_F=-1.0$ eV, consistent with previous findings\cite{Harfah2022}. However, in the lower Ni slab, hybridized with hBN(V$_B$), the DOS peak in the spin minority channel at $E-E_F=0.4$ eV is reduced due to the electron localization at the vacancy site. A similar reduction is observed in the spin majority channel around $E-E_F=-1.0$ eV. The disparity is more pronounced between the upper and lower hBN layers, as shown in Figure \ref{fig:3rd_figure1}(c). In the upper hBN, a nearly insulating gap dominates the DOS, with a peak only appearing at $E-E_F=0.4$ eV for the spin minority channel and around $E-E_F=-1.0$ eV for the spin majority channel, arising from $pd$-hybridization. Conversely, the lower hBN containing V$_B$ exhibits a DOS peak in the same energy range as the upper hBN, along with additional states. In the spin majority channel, a higher DOS peak is observed from $E-E_F=-0.9$ to $-0.2$ eV, with the highest and second-highest DOS at $E-E_F=-0.38$ and $-0.65$ eV, respectively. Meanwhile, in the spin minority channel, a broad range of new states emerges from $E-E_F=-1.0$ to $1.0$ eV. The higher occupancy of states in the spin majority channel compared to the minority channel corresponds to the spin-up magnetization of hBN(V$_B$) due to the vacancy.

\begin{figure}[tb]
\centering
\includegraphics[width=\columnwidth]{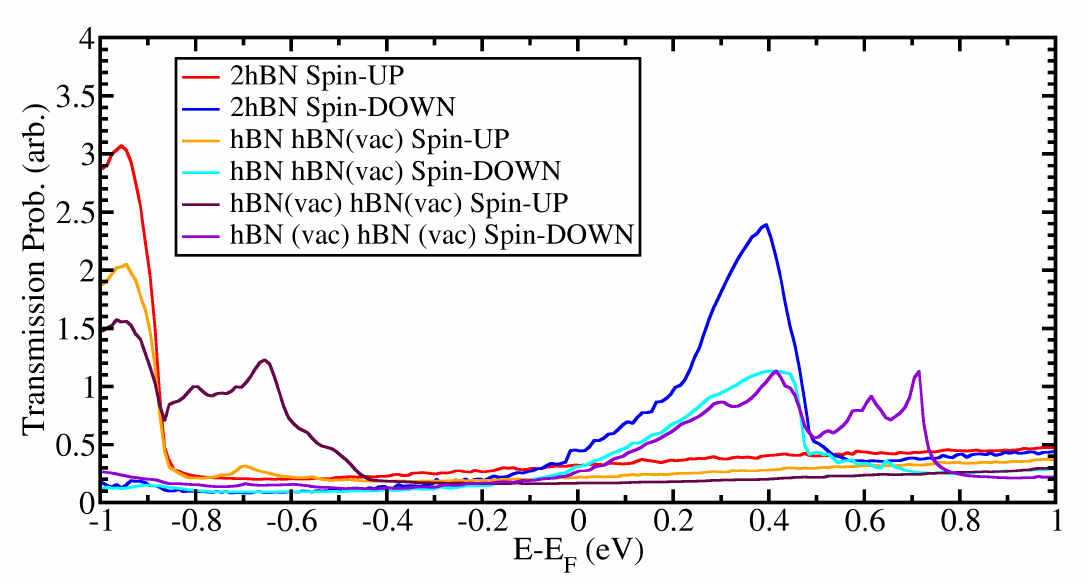}
\caption{\label{fig:3rd_figure3} Comparison of transmission probability for Ni/2hBN/Ni, Ni/hBN(V$_B$)/hBN/Ni, and Ni/hBN(V$_B$)/hBN(V$_B$)/Ni MTJs.}
\end{figure}

The modification of the LDOS in the lower Ni and hBN layers directly affects the probability of electron transmission. As previously mentioned, in a two-layer hBN MTJ system, electron transmission primarily occurs through the Ni surface state. Figure \ref{fig:3rd_figure3} illustrates a prominent transmission peak at $E-E_F=0.38$ eV ($-0.98$ eV) for spin-down (spin-up) electrons, consistent with previous findings. This high transmission is attributed to electron transmission through the Ni d$_{z^2}$ orbital surface state. However, in the Ni/hBN/hBN(V$B$)/Ni MTJ structure, the decreased Ni d${z^2}$ orbital DOS leads to a significant reduction in electron transmission for both spin-up and spin-down electrons. Conversely, in the Ni/hBN(V$_B$)-hBN(V$_B$)/Ni MTJ, identical LDOS in both the upper and lower Ni and hBN layers due to the creation of vacancies on both hBN layers, as shown in Figure \ref{fig:3rd_figure2}(b) and (c), similarly reduces electron transmission probability for spin-down (spin-up) electrons at $E-E_F=0.38$ eV ($-0.98$ eV) while creating a new transmission channel for spin-down (spin-up) electrons from $E-E_F=0.5$ eV ($-0.85$ eV) to $0.8$ eV ($-0.4$ eV). This suggests that the newly created states in hBN(V$B$) mediate electron transmission not only for the Ni d${z^2}$ orbital surface state but also for another orbital surface state.

We note that the Ni/hBN(V$_B$)-hBN(V$_B$)/Ni MTJ introduces a new transmission channel when the hBN(V$_B$) layers are in close proximity. However, as the separation between hBN(V$_B$) layers increases, the intensity of the transmission channel decreases and eventually disappears, resembling the transmission observed in Ni/hBN(V$_B$)-hBN/Ni.

\subsection{Interaction Between Ni's Surface State at Interface with Localized State of 2D Materials Vacancy}

In the preceding section, we examined the influence of V$_B$ on the electronic structure of both the hBN and Ni layers at the Ni/hBN interface. Here, we delve into the interaction between the surface state of Ni at the interface, which remains unaltered, and the isolated localized state of the middle hBN layer in the Ni/hBN/hBN(V$_B$)/hBN/Ni MTJ.

\begin{figure}[tb]
\centering
\includegraphics[width=\columnwidth]{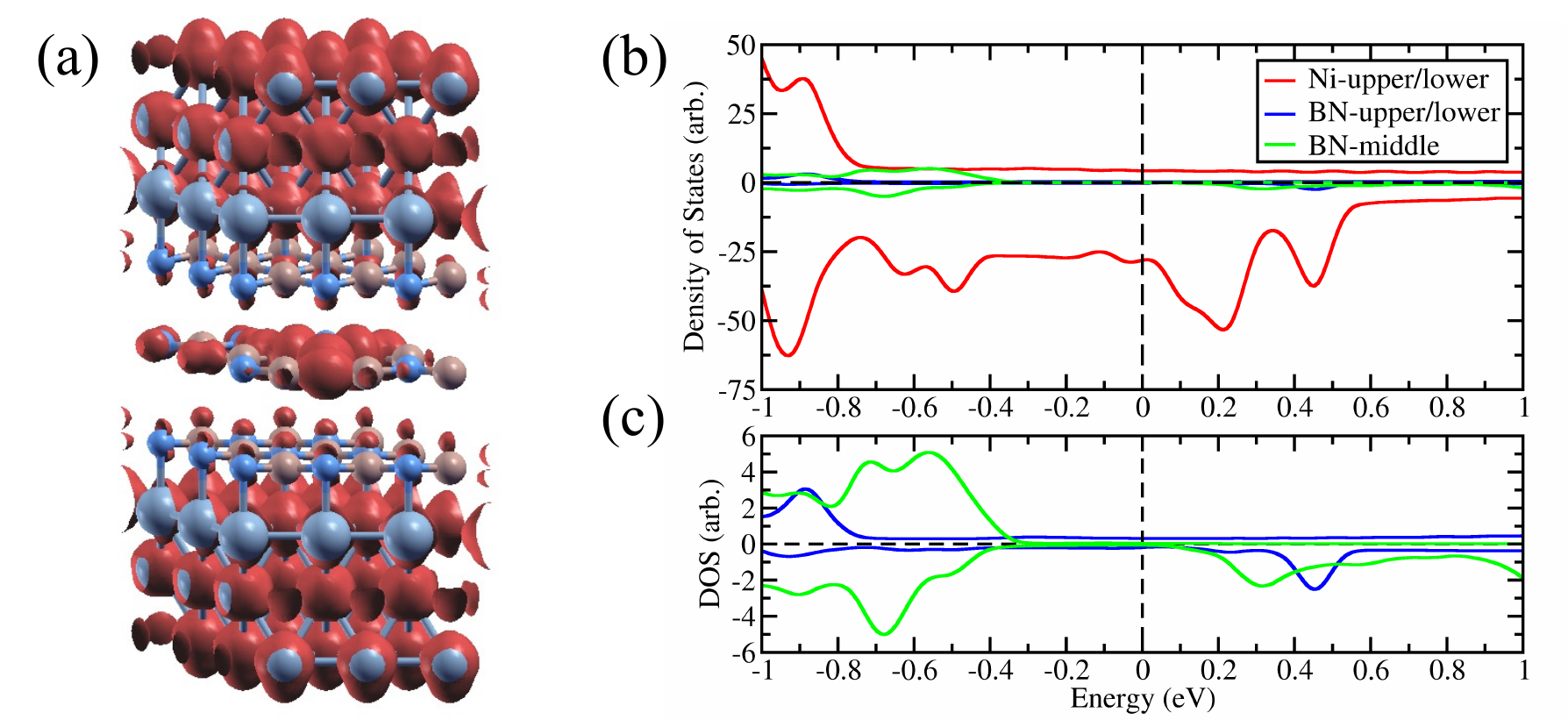}
\caption{\label{fig:3rd_figure4} (a) The $3 \times 3$ Ni/hBN/hBN(V$_B$)/hBN/Ni SCDM (red color represents spin-up electron density). Visualization was performed using XCrySDen \cite{xcrysden}. (b) Local density of states (LDOS) for Ni-upper/lower and hBN-upper/lower (hBN without V$_B$), and hBN-middle (hBN with V$_B$) in PC configuration. The positive (negative) value of DOS represents spin majority (minority) channel. (c) The zoom on lower energy of LDOS in (b).}
\end{figure}

Figure \ref{fig:3rd_figure4}(a) illustrates the spin charge density mapping (SCDM) of the Ni/hBN/hBN(V$_B$)/hBN/Ni MTJ, indicating that the hBN(V$_B$) in the middle possesses an induced magnetic moment. A strong magnetic moment is observed at the vacancy site, aligned parallel to the Ni slab, suggesting magnetic interaction between the hBN(V$_B$) and Ni slab despite lacking a direct interface. The local density of states (LDOS) of the Ni slab, hBN at the interface, and hBN(V$_B$) is depicted in Figures \ref{fig:3rd_figure4}(b) and (c). The LDOS analysis reveals that the hBN(V$_B$) in the middle does not alter the DOS of the Ni layer at the interface or the hBN at the interface. However, the DOS of the hBN(V$B$) exhibits spin polarization, evident from the Stoner gap. The newly created state from the vacancy shows some overlap with the surface state of the Ni d${z^2}$ orbital at $E-E_F=0.42$ eV ($-0.9$ eV) for the spin minority (majority) channel.

\begin{figure}[tb]
\centering
\includegraphics[width=\columnwidth]{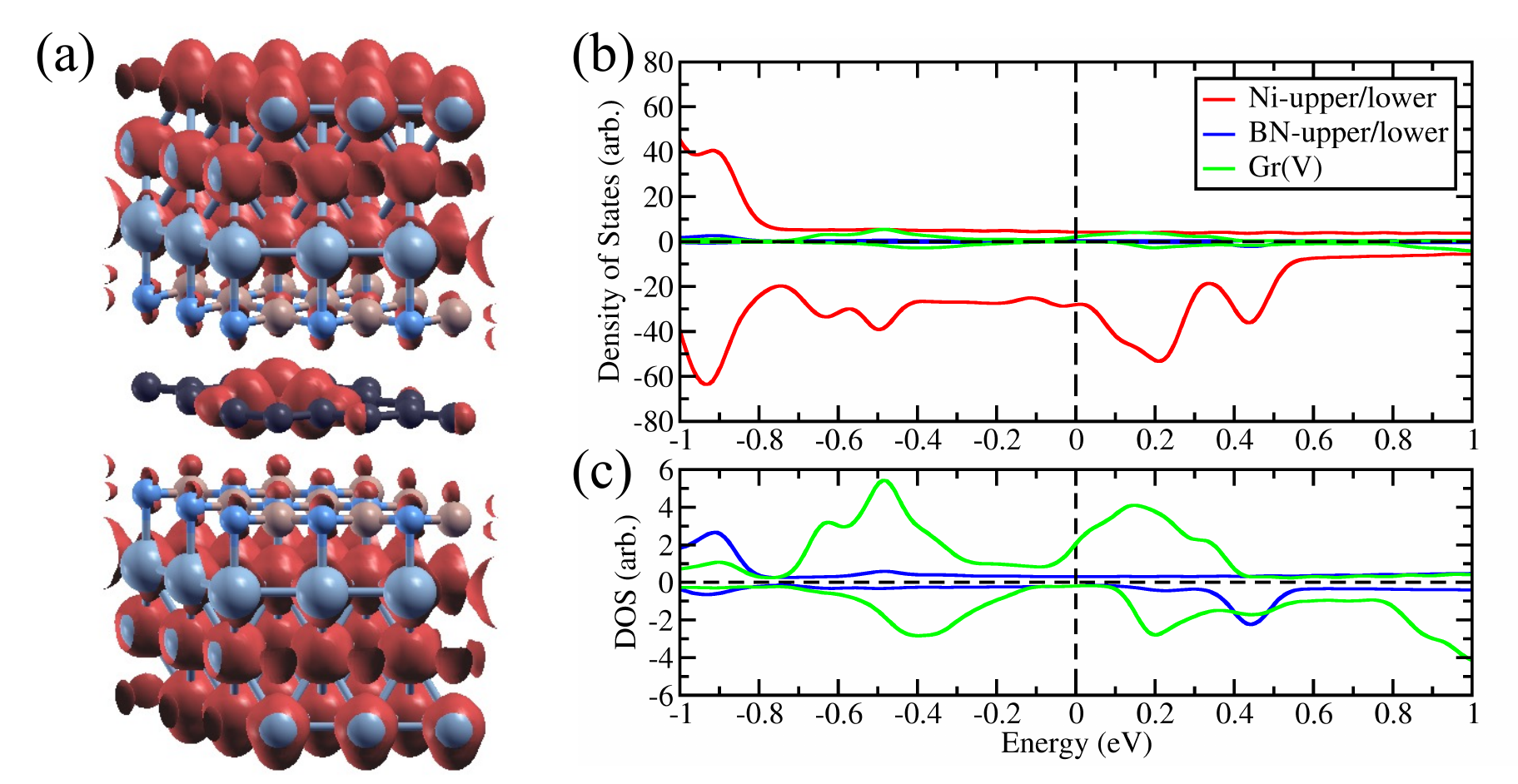}
\caption{\label{fig:3rd_figure5} (a) The $3 \times 3$ Ni/hBN/Gr(V)/hBN/Ni SCDM (red color represents spin-up electron density). Visualization was performed using XCrySDen \cite{xcrysden}. (b) LDOS for Ni-upper/lower, hBN-upper/lower, and Gr(V) in PC configuration. The positive (negative) value of DOS represents spin majority (minority) channel. (c) The zoom on lower energy of LDOS in (b).}
\end{figure}

\begin{figure}[tb]
\centering
\includegraphics[width=\columnwidth]{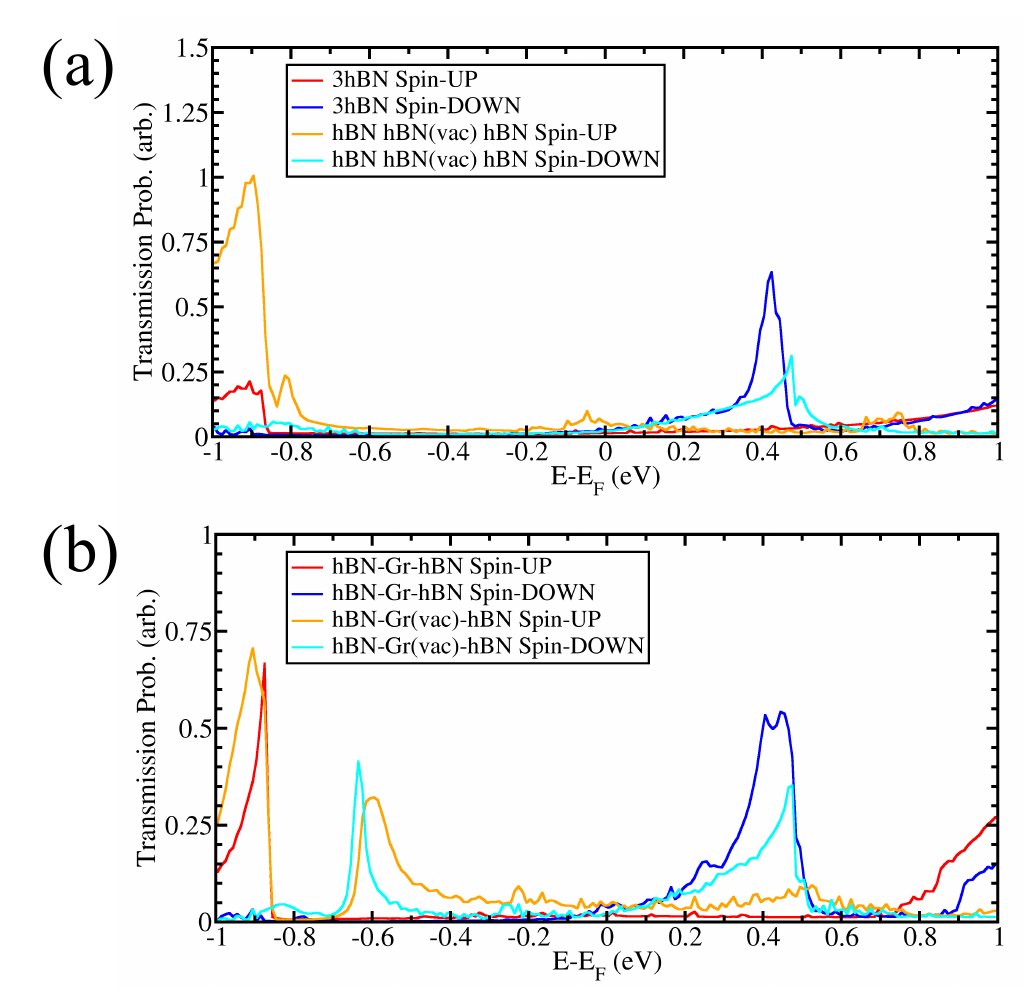}
\caption{\label{fig:3rd_figure6} Transmission probability comparison between (a) Ni/3hBN/Ni and Ni/hBN/hBN(V$_B$)/hBN/Ni MTJs, and between (b) Ni/hBN/Gr/hBN/Ni and  Ni/hBN/Gr(V)/hBN/Ni MTJs.}
\end{figure}

The transmission probability of the Ni/hBN/hBN(V$B$)/hBN/Ni MTJ is displayed in Figure \ref{fig:3rd_figure6}(a). No new transmission channel is observed. Instead, an increment (reduction) in the transmission probability is noted for the spin majority (minority) channel at $E-E_F=-0.9$ eV ($0.42$ eV) compared to the Ni/3hBN/Ni MTJ. This indicates that the Ni d${z^2}$ orbital solely affects the occupied state of the localized state. The reduction in transmission probability in the spin minority channel, due to the Ni d${z^2}$ orbital surface state, does not interact with the proximity effect, which exhibits spin-up electron localization. Conversely, the spin majority channel of the Ni d${z^2}$ orbital surface state propagates through the localized state, resulting in an increment in the transmission probability of spin-up electrons.

In a complementary investigation, we introduced a monoatomic vacancy in the graphene layer to create the Ni/hBN/Gr(V)/hBN/Ni MTJ. Similar to the hBN(V$_B$) in Ni/hBN/hBN(V$_B$)/hBN/Ni MTJ, the graphene with a vacancy exhibits an induced magnetic moment and ferromagnetic interaction with the Ni slab, as shown in Figure \ref{fig:3rd_figure5}(a). Furthermore, the vacancy in graphene does not modify the DOS of the Ni layer at the interface or the hBN layer at the interface, as depicted in Figures \ref{fig:3rd_figure5}(b) and (d). Interestingly, a decrease in spin-down electron transmission at higher energy, akin to the Ni/hBN/hBN(V$_B$)/hBN/Ni MTJ system, is observed. In the Ni/hBN/Gr(V)/hBN/Ni MTJ, the vacancy in graphene creates a new transmission channel for both spin-up and spin-down electrons within the same energy range, around $E-E_F=-0.6$ eV, as shown in Figure \ref{fig:3rd_figure6}(b). This unique transmission channel originates from the proximity effect, which interacts with the localized state as well as the $\pi$-orbital of the graphene layer. However, this interaction occurs at the occupied state, implying that the surface state not only propagates through Gr(V) mediated by the localized state but can also be redistributed within the graphene layer through graphene's $\pi$-orbital. We anticipate the tunneling magnetoresistance (TMR) ratio at $E-E_F=-0.6$ eV to exceed 1200\%, as found in Ni/hBN/Gr/hBN/Ni.

\section{Conclusion}
In this study, we investigated the impact of a monoatomic vacancy in 2D materials on the performance of an MTJ device. Initially, we explored bilayer hexagonal boron nitride (hBN) configurations, considering Ni/hBN(V$_B$)-hBN/Ni and Ni/hBN(V$_B$)-hBN(V$_B$)/Ni MTJ systems. The transmission probability was notably reduced in the Ni/hBN(V$_B$)-hBN/Ni configuration compared to the pristine bilayer hBN system. Conversely, in the Ni/hBN(V$_B$)-hBN(V$_B$)/Ni configuration, an additional peak in the transmission probability emerged, indicating a new transmission channel facilitated by the localized state of hBN(V$_B$)-hBN(V$_B$).

Subsequently, we introduced a monoatomic boron vacancy in the middle hBN layer of the Ni/3hBN/Ni system, creating a Ni/hBN-hBN(V$_B$)-hBN/Ni MTJ. The presence of the localized state in the middle hBN layer led to a reduction (enhancement) in the transmission probability peak for the spin minority (majority) channel at higher (lower) energy, attributed to the stoner gap created in the middle hBN layer. Consequently, the transmission probability for the spin-majority channel was augmented.

Finally, we investigated the monoatomic vacancy in the graphene layer of the Ni/hBN-Gr-hBN/Ni MTJ. The presence of the graphene vacancy induced a new transmission channel for both the spin-majority and minority channels within the same energy range. This unique transmission channel originated from the proximity effect interacting with the localized state of the graphene layer.

The introduction of a monoatomic vacancy on the insulator barrier of 2D materials in the MTJ system results in unique characteristics determined by the interaction between the surface state of the electrode and the localized state of the monoatomic vacancy layer. This interaction offers promising avenues for tailoring the properties and performance of spintronic devices based on MTJs.

\section*{Conflicts of interest}
``There are no conflicts to declare''.

\section*{Acknowledgements}
Calculations were performed at the Kyushu University computer center. H.H. gratefully acknowledges the fellowship support from the JSPS. This study was partly supported by the Japan Society for the Promotion of Science (JSPS) KAKENHI (Grant Nos. JP19H00862 and JP16H00914 in the Science of Atomic Layers, 21J22520 in the Grant-in-Aid for Young Scientists, and JP18K03456).

\bibliography{rsc}

\end{document}